\documentclass{article}

\usepackage{arxiv}

\usepackage[utf8]{inputenc} % allow utf-8 input
\usepackage[T1]{fontenc}    % use 8-bit T1 fonts
\usepackage{url}            % simple URL typesetting
\usepackage{booktabs}       % professional-quality tables
\usepackage{amsfonts}       % blackboard math symbols
\usepackage{nicefrac}       % compact symbols for 1/2, etc.
\usepackage{microtype}      % microtypography
\usepackage{lipsum}

\usepackage{relsize,balance,lipsum,bbm,enumerate,times,comment,color,graphicx,setspace,mathdots,mathrsfs,amssymb,latexsym,amsfonts,amsmath,cite,stmaryrd,caption,pgf,tikz,accents,mathtools,tabu,enumitem,hhline,array,epstopdf,nicefrac,amsthm,microtype,algorithmic,array,float,bm,url}

\usepackage{graphicx}
\graphicspath{ {./images/} }

\usepackage{relsize,balance,lipsum,bbm,enumerate,times,comment,color,graphicx,setspace,bbm,mathdots,mathrsfs,amssymb,latexsym,amsfonts,amsmath,cite,stmaryrd,caption,pgf,tikz,accents,mathtools,tabu,enumitem,hhline,array,epstopdf}
\usepackage[ruled,vlined]{algorithm2e}
\usepackage{multirow}
\usepackage[utf8]{inputenc}
\graphicspath{{./images/}}
\usepackage[font=small,skip=0pt]{caption}
\usepackage{amssymb,multicol}
\usepackage{enumerate}
\usepackage{dirtytalk}
\usepackage{makecell}
\allowdisplaybreaks
\usepackage{amsthm}

\usepackage{amsthm}

\title{Incorporating Coincidental Water Data into Non-intrusive Load Monitoring}

\author{
    Mohammad-Mehdi~Keramati$^{\dagger}$\\
    Dept. of Electrical and Computer Engineering\\
    Tarbiat Modares University\\
    Tehran, Iran\\
    \texttt{m.keramati@modares.ac.ir}
\And    
    Elnaz~Azizi$^{\dagger}$\\
    Dept. of Electrical and Computer Engineering\\
    Tarbiat Modares University\\
    Tehran, Iran\\
    \texttt{e.azizi@modares.ac.ir}
\And
    Hamidreza~Mohemi\\
    Dept. of Electrical and Computer Engineering\\
    Tarbiat Modares University\\
    Tehran, Iran\\
    \texttt{momeni\_h@modares.ac.ir}
\And
    Sadegh~Bolouki\\
    Dept. of Electrical and Computer Engineering\\
    Tarbiat Modares University\\
    Tehran, Iran\\
    \texttt{bolouki@modares.ac.ir}
\And
$^{\dagger}$Equal Contribution.
}

\theoremstyle{definition}

\begin{document}
\maketitle

\begin{abstract}
    Non-intrusive load monitoring (NILM) as the process of extracting the usage pattern of appliances from the aggregated power signal is among successful approaches aiding residential energy management. In recent years, high volume datasets on power profiles have become available, which has helped make classification methods employed for the NILM purpose more effective and more accurate. However, the presence of multi-mode appliances and appliances with close power values have remained influential in worsening the computational complexity and diminishing the accuracy of these algorithms. To tackle these challenges, we propose an event-based classification process, in the first phase of which the $K$-nearest neighbors method, as a fast classification technique, is employed to extract power signals of appliances with exclusive non-overlapping power values. Then, two deep learning models, which consider the water consumption of some appliances as a novel signature in the network, are utilized to distinguish between appliances with overlapping power values. In addition to power disaggregation, the proposed process as well extracts the water consumption profiles of specific appliances. To illustrate the proposed process and validate its efficiency, seven appliances of the AMPds are considered, with the numerical classification results showing marked improvement with respect to the existing classification-based NILM techniques.
\end{abstract}
    

%%%%%%%%%%%%%%%%%%
%%%%%%%%%%%%%%%%%%
\section{Introduction}

    In recent years, residential demand-side management has become an effective policy in smart grids to achieve objectives such as reducing energy consumption and consumers' costs. Studies show that non-intrusive load monitoring (NILM), as a tool aiding demand-side management, can be efficient in reducing energy consumption and the subsequent energy costs up to 15\% in the residential sector \cite{darby2006effectiveness}. NILM is the process that infers the consumption profile of each appliance of a residence given the aggregated power signal by purely analytical algorithms. The information extracted by NILM can be used as feedback to both consumers and providers to improve the efficiency of the entire grid. From the consumers' standpoint, detailed information with regard to the usage pattern of each appliance helps detect faulty appliances. It also shows how the consumers should change their usage behavior and promotes a rational consumption to reduce electricity costs and save energy while maintaining their comfort \cite{welikala2019incorporating}. From the providers' standpoint, the detailed consumption pattern and behavior of consumers can be used to adopt efficient strategies to strike the balance between demand and generation and overcome various grid's challenges \cite{yang2020event}. 
 
%The simplest load monitoring method, called intrusive load monitoring, is installing sensors on all appliances and monitoring their usage pattern. However, it is an expensive and invasive approach. Therefore, the non-intrusive load monitoring (NILM) technique, in which the usage patterns of appliances are extracted from the given aggregated consumption signal by purely analytical methods, gains a lot of attention in recent years.

    The concept of NILM was originally introduced by G.W.~Hart in 1992 \cite{hart1992nonintrusive}. In recent years, different algorithms have been proposed for the NILM problem. These algorithms can be divided into two main categories, learning-based approaches (supervised, semi-supervised and unsupervised methods) and optimization-based methods. A comprehensive review of these methods can be found in \cite{esa2016review}. Due to the presence of high volume training datasets in recent years, the majority of algorithms have been based on various data-driven learning-based techniques, depending on the sampling rate of the signal and features of appliances. Most researchers in this field have been interested in employing low-sampling load measurements since they require inexpensive sensors as well as low storage memories \cite{zeifman2011nonintrusive}. Furthermore, these data are obtainable from the existing smart meters \cite{xu2018classifier}. From another viewpoint, due to the low computational complexity of event-based methods, which are based on the mode transitions of appliances, including their ON/OFF transitions, event-based approaches have become increasingly popular in recent years \cite{he2019generic}.

    The presence of multi-mode appliances and appliances with close power values increase the computational complexity and often reduce the accuracy of event-based NILM methods \cite{8720065}. This paper aims to address these challenges by combining two classification methods, namely KNN and LSTM, where the former method is known for its speed and the latter for its accuracy. Furthermore, due to the correlation between the electricity and water usage of some appliances, water consumption is considered as a new feature in the network to increase the accuracy of load monitoring.

%%%%%%%%%%%%%%%%%%
\subsection{Related Work}

    Event-based NILM algorithms are classified into two major categories, supervised and unsupervised. Supervised algorithms require ground-truth information about appliances to be trained \cite{dong2012event}. Unlike supervised NILM algorithms, unsupervised methods extract different clusters of appliances and do not require a training dataset. In comparison with the supervised algorithms, although unsupervised ones are less costly, they generally have lower accuracy \cite{anderson2014non}. Therefore, supervised methods, and particularly classification algorithms, have gained a great deal of attention in the NILM research. Classification algorithms extract the usage pattern of each appliance from the aggregated power signal based on a pre-learned model. Various classification methods such as KNN (K-nearest neighbor) \cite{schirmer2019statistical}, support vector machines (SVM) \cite{singh2015residential}, multi-label classification \cite{tabatabaei2017toward}, and decision tree \cite{gillis2015nonintrusive}, have been proposed for the NILM problem.
    
    Due to the emergence of high volumes of data in recent years, hence the capability of learning hidden features, deep neural networks have returned relatively accurate results for the NILM purpose. In the first exploration, Kelly and Knottenbelt in \cite{kelly2015neural} evaluated the facility of adopting neural networks to address the NILM problem via three deep neural network architectures. The method of sequence-to-point was suggested for using convolutional neural networks (CNN) in \cite{zhang2018sequence}, while \cite{de2018appliance} offered V-I trajectories as a CNN architecture input to classify appliances. Also, a non-uniform sub-sampling scheme considering the auto-encoder network for energy disaggregation was presented in \cite{fagiani2019non}, while authors in \cite{mauch2015new} suggested a deep recurrent, long short-term memory (LSTM) network to carry out NILM. In \cite{chavat2020non}, an adopted LSTM architecture and auto-encoder networks were proposed to detect similarities in consumption patterns. %The high computation time and complexity are the main challenge of the majority of the deep learning-based NILM methods.

    The main drawback of the classification-based methods is their inefficiency in distinguishing appliances with close power values from each other. To tackle this issue, additional features and signatures of appliances such as light and motion have been utilized in \cite{srinivasan2013fixturefinder}. Time and ON/OFF duration of appliances were considered in \cite{elbe2013appliance}. The exterior temperature was proposed in \cite{wytock2014contextually}, while appliances' noise during their time of usage was suggested in \cite{patel2007flick} as  extra features in load disaggregation. However, utilizing these additional features require additional sensors that, besides the high costs, is intrusive and can disturb residents.% Extracting other features requires more manual effort which is not practical.

    One of the specific features of appliances, which can be utilized in NILM for separating appliances with close power values, is their simultaneous water or gas consumption. Utilizing the power profiles for water disaggregation has been proposed in different studies \cite{cominola2015benefits, ellert2016appliance, nguyen2017water}. Most notably, authors in \cite{ellert2016appliance} adopted power as a novel signature in the water disaggregation issue, while a combination of hidden Markov model, deep learning, and energy to obtain higher accuracy in water disaggregation problems was proposed in \cite{nguyen2017water}. The elaboration of the last method will dwindle by using LSTM. However, to the best of the authors' knowledge, this is the first time that water consumption of appliances is utilized to improve the load disaggregation accuracy.

%As well as load monitoring, there are various benefits for water end-use disaggregation such as preventing water theft, leakage detection, and client service improvement.\cite{hawkins2015considerations} A similar blind identification problem like NILM uses a pattern of fixture water consumption to classify whole-home water consumption.\cite{cominola2015benefits} Identi flow, HydroSense and Autoflow are famous tools in the field of non-intrusive water consumption monitoring. Identiflow by employing a decision tree algorithm for water disaggregation consumption semi-autonomously was introduced by Kowalski and Marshallsay.\cite{kowalski2003system} Froehlich et al.\cite{froehlich2009hydrosense} proposed a Bayesian approach for identifying pressure waves of individual appliances in HydroSense software. Nguyen et al.\cite{nguyen2015intelligent} developed autoflow by using machine learning algorithms such as the hidden Markov model, artificial neural networks, and etc. Although proper accuracy of water disaggregation has been shown in the studies mentioned, considering other aspects of energy consumption of residential consumers such as electricity and gas can reduce the complexity of the disaggregation process.

%%%%%%%%%%%%%%%%%%
\subsection{Contributions}

    Even though event-based NILM methods are popular due to their low complexity, the presence of appliances with close power values has become a striking challenge in this line of research. Deep learning-based methods have proved to be fairly accurate in tackling this issue due to their capability in feature learning. However, these algorithms are more complex. Keeping these challenges in mind, the key contributions of this paper can be summarized as follows:

    \begin{enumerate}
        \item Instead of utilizing only a fast classification method or only an accurate deep-learning neural network, we propose a paradigm that takes advantage of both of these method types. More specifically, the proposed technique benefits from both the low computational complexity of KNN and the high accuracy of LSTM.
        \item Unlike other studies that consider only the active power of appliances as their specific signature, we as well use the water consumption of some appliances {\color{black}(only the dishwasher in our case study)} as a novel appliance signature for the network, which leads to an increase in the accuracy of NILM.
        \item In addition to extracting the power consumption profiles of appliances, the proposed method also extracts the water consumption profiles of specific appliances {\color{black}(the dishwasher in our case study)} accurately.
        %\item The proposed method performs well given low sampling data, which is more practical in the real world.
    \end{enumerate}

%%%%%%%%%%%%%%%%%%
\subsection{Paper Organization}

    The remainder of the paper is organized as follows. In Section II, some fundamental concepts such as event detection, KNN classification, LSTM network, and class balancing are briefly discussed and the problem is formally presented. The pre-processing phase is detailed in Section III. Our classification-based techniques for the NILM problem are proposed and explained in Section IV. Simulations and numerical experiments are demonstrated in Section V. Section VI compares the efficiency of the proposed methods given various circumstances. Finally, Section VII concludes the paper.

%%%%%%%%%%%%%%%%%%
%%%%%%%%%%%%%%%%%%
\section{Fundamental Concepts and Problem Statement}

    In this section, we discuss the fundamental concepts and tools used in this paper and formulate the problem.

%%%%%%%%%%%%%%%%%%
\subsection{Event Detection}

    A fundamental step in any event-based NILM method is detecting the events accurately, where an event in general refers to a mode transition of any appliance. Due to the presence of noise and fluctuations in the power signal, it is essential to establish a reliable event detection procedure. In a majority of research articles on event-based NILM, an event is detected based on the difference of two consecutive samples considering a predefined threshold. Determining a proper threshold to not miss any actual event or to not mistake overshoots or transient spikes for events is challenging and a topic of active research. This paper relies on the method proposed in {\color{black}the authors' earlier work} \cite{azizi2020residential} for event detection, where events are detected based on the local mean and standard deviation of three consecutive samples. This method consists of three main steps briefly explained below.

    \noindent \textbf{Step 1:} From the original signal $P(t)$, events of which are to be detected, a simple moving average (SMA) is constructed based on every three consecutive samples as 
	\begin{equation}
		\bar{P}(t) = \frac{1}{3}\sum_{\tau=0}^2 P(t-\tau)
	\end{equation}
    
    \noindent\textbf{Step 2:} The standard deviation $\sigma_P$ of every three consecutive samples of $P(t)$ is calculated and a piece-wise constant signal $P_S(t)$ is constructed according to
	\begin{equation}\label{eq2}
		\sigma_P^2(t) = \frac{1}{3} \sum_{\tau=1}^3 \left( P(t-\tau) - \bar{P}(t) \right)^2
	\end{equation}
	\begin{equation} \label{eq3}
	\begin{array}{cc}
		\text{if } \sigma_P(t) < \sigma_g:& ~ P_S(t) = \bar{P}(t)\\
		\text{if } \sigma_P(t) \geq \sigma_g:& ~ P_S(t) = P_S(t-1)
    \end{array}
	\end{equation}
    where $\sigma_g$ is the standard deviation of the power grid’s noise.
    
    \noindent\textbf{Step 3:} Events are detected by calculating the difference of two consecutive samples of $P_S(t)$ and comparing it to $\sigma_g$, i.e.,
	\begin{equation} \label{eq4}
		\Delta P_S(t) = P_S(t) - P_S(t-1)
	\end{equation}
	\begin{equation} \label{eq5}
		\text{if } \Delta P_S(t) \geq \sigma_g:~ T(q) = t,~ {\color{black}q=q+1,}
	\end{equation}
    where $T(q)$ indicates the time of the $q$th event. 

%%%%%%%%%%%%%%%%%%
\subsection{KNN Classification Method}

%Multi-label classification was first introduced in the text categorization problem. Indeed each document may involve several different topics simultaneously. Consider data set D with input X and number of samples n , which  Y  represents the label of each sample $(X_1,Y_1 ),(X_2,Y_2 ),(X_3,Y_3 ),…,(X_n,Y_n )$. If the number of labels (Y) that assigns for each sample be more than two, it is called multi-label classification.
%Methods for multi-label classification divided into two categories problem, transformation, and algorithm adaption. Transformation methods such as label power set, binary relevance, etc tackle multi-label classification problems by transforming it into single-label classification. Algorithm adaptation techniques such as ML-KNN, ML-DT, etc address multi-label classification problems by using adapting popular learning methods like KNN, DT, etc to handle multi-label datasets directly. Due to the simplicity and low computation burden, MLKNN will be utilized in the first phase of this paper \cite{tabatabaei2016toward}.

    Classification algorithms are models that learn from a training dataset to identify to which class new data points belong. The simplest classification algorithm, which is also the most commonly used due to its simplicity and low computational complexity, is the $K$-nearest neighbor (KNN) algorithm. %This algorithm has three critical defining elements, 1) distance measurement (e.g. Manhattan, Minkowski, etc), 2) $K$, which is the number of neighbors, and 3) classification determination rules.
    Incoming a new data point, KNN classifies it where a majority of its closest $K$ data points in the training dataset belongs \cite{guo2003knn}. In implementing KNN, the number $K$, which {\color{black}must be greater than 1}, is a hyperparameter and should be tuned. Moreover, a proper metric quantifying the distance between data points should be adopted.
%\subsubsection{MLKNN} Multi-label KNN method act as a helping hand to disaggregate total power consumption besides finding overlapping appliances' power consumption profiles. In this approach, a classifier for each appliance had been trained separately after the events of total power consumption signal were detected. Event i of the total power consumption signal that was caused by appliance j must have just a label 1 at each sample. Otherwise, overlapped happened \cite{tabatabaei2016toward}. 

%%%%%%%%%%%%%%%%%%
\subsection{LSTM Network}

\begin{figure}[t]
        \centering
        \includegraphics[width=.5\linewidth]{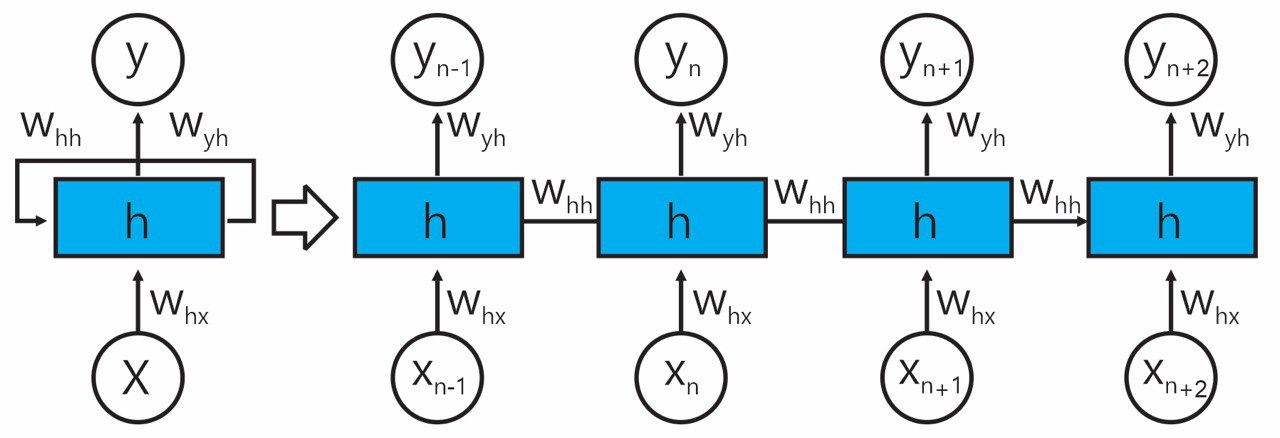}\vspace{.1in}
        \caption{RNN architecture}
        \label{rnn}
    \end{figure}

    A popular neural network model for the classification of time-series data is the recurrent neural network (RNN). The RNN differs from traditional feedforward neural networks in that it has a feedback connection, as shown in Fig.~\ref{rnn}, which establishes a connection between previous information and the current one. This feedback makes the RNN effective for time-series data classification and regression \cite{kelly2015neural}. The input in the NILM problem is the power signal, which is of the time-series type, suggesting that the RNN model could be suitable. However, back-propagation training with these methods often fails because of vanishing/exploding gradients. Among the most well-known and effective RNN models is the long short-term memory (LSTM) model, which overcomes these challenges because of its ability to store information for a long period of time.

%%%%%%%%%%%%%%%%%%
\subsection{Class Balancing}

    As in the case of numerous well-known problems in which the challenge of unbalanced datasets exist, such as cancer classification and fraud detection, there are some classes with significantly more samples than other ones in any NILM dataset. To achieve class balance, resampling methods have been employed. Resampling may be performed via under-sampling, where some randomly chosen samples are removed, or over-sampling that multiplies some randomly chosen samples in the dataset. While over-sampling is often more effective than under-sampling for class balancing, it leads to a drastically larger dataset, processing of which is more complex. Considering this trade-off, an under-sampling method, proposed in \cite{fang2020non}, is selected to carry out class balancing in this work. 

%%%%%%%%%%%%%%%%%%
\subsection{Problem Statement}

    As mentioned previously, NILM is the process of extracting the consumption profile of each appliance given the aggregated power signal via analytical algorithms. Event-based NILM classification can be defined as assigning appliances, as classes or labels, to each event of the aggregated signal. In other words, if a mode transition of appliance $i$ caused an event in the aggregated signal, that event should get label $i$. Due to the fact that some appliances, such as the dishwasher, also consumes water, there are correlations between the aggregated power and water consumption signals of a residence. We aim to take advantage of these correlations to design an accurate event-based NILM algorithm.

\section{Data Pre-processing}\label{preprocessing}

    The AMPds consists of time-series data for the power signals of various appliances and the water signals of some appliances, such as the dishwasher. Based on these data, the aggregated power and water signals of the house can be obtained. Then, in the data pre-processing stage, the following tasks are completed to obtain the necessary data for training the NILM model.% First, events of power signals of individual appliances, along with those of the aggregated signal, are detected. Afterward, a signal is constructed based on the events of the aggregated power signal. Finally, proper labels are assigned to the events of the constructed signal.

%%%%%%%%%%%%%%%%%%
\begin{enumerate}
    \item {\bf Constructing the event signal.} After applying the event detection method borrowed from \cite{azizi2020residential} to the aggregated power signal and obtaining the filtered signal $P_S(t)$, its corresponding {\it event signal}, which has the same domain as the aggregated power signal, is constructed as follows. The value of a sample in the event signal is 0 unless an event has occurred at that time in the aggregated signal, in which case the sample gets a value equal to the event value, defined as the change in the power value that occurred during the event. One notices that the event signal in a way resembles a finite difference of the aggregated signal. Fig.~\ref{signal_events} illustrates the aggregated power signal and its corresponding event signal for some time period.
    \begin{figure}[t]
        \centering
        \includegraphics[width=.5\linewidth]{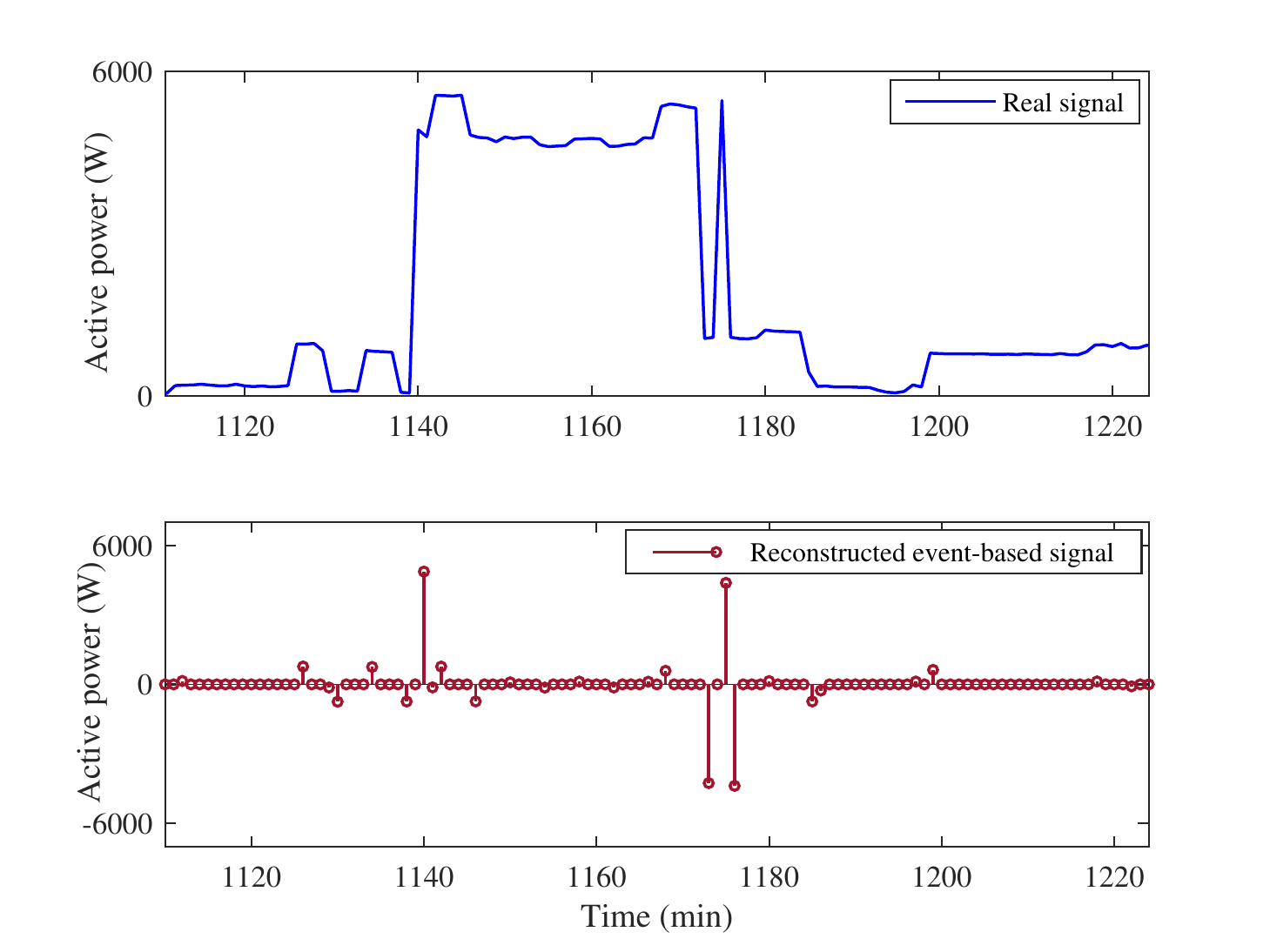}\vspace{.05in}
        \caption{The power signal and its corresponding event signal}
        \label{signal_events}
    \end{figure}

    \item {\bf Event labeling for the power signal.} Given an event of the aggregated power signal {\color{black}in the training dataset}, or equivalently a non-zero sample in the event signal, it is labeled $a_{ij}$ if it is caused by {\color{black}mode transition} $j$ of appliance $i$.

    \item {\bf Event labeling for the water signal.} Given the training dataset, a sample of the water signal of an appliance is labeled 1 if {\color{black}the sampled value} is non-zero. Otherwise, it is labeled 0. Fig.~\ref{water_recons} illustrates labels corresponding to the water signal of the dishwasher for some time period.
    \begin{figure}[t]
        \centering
        \includegraphics[width=.5\linewidth]{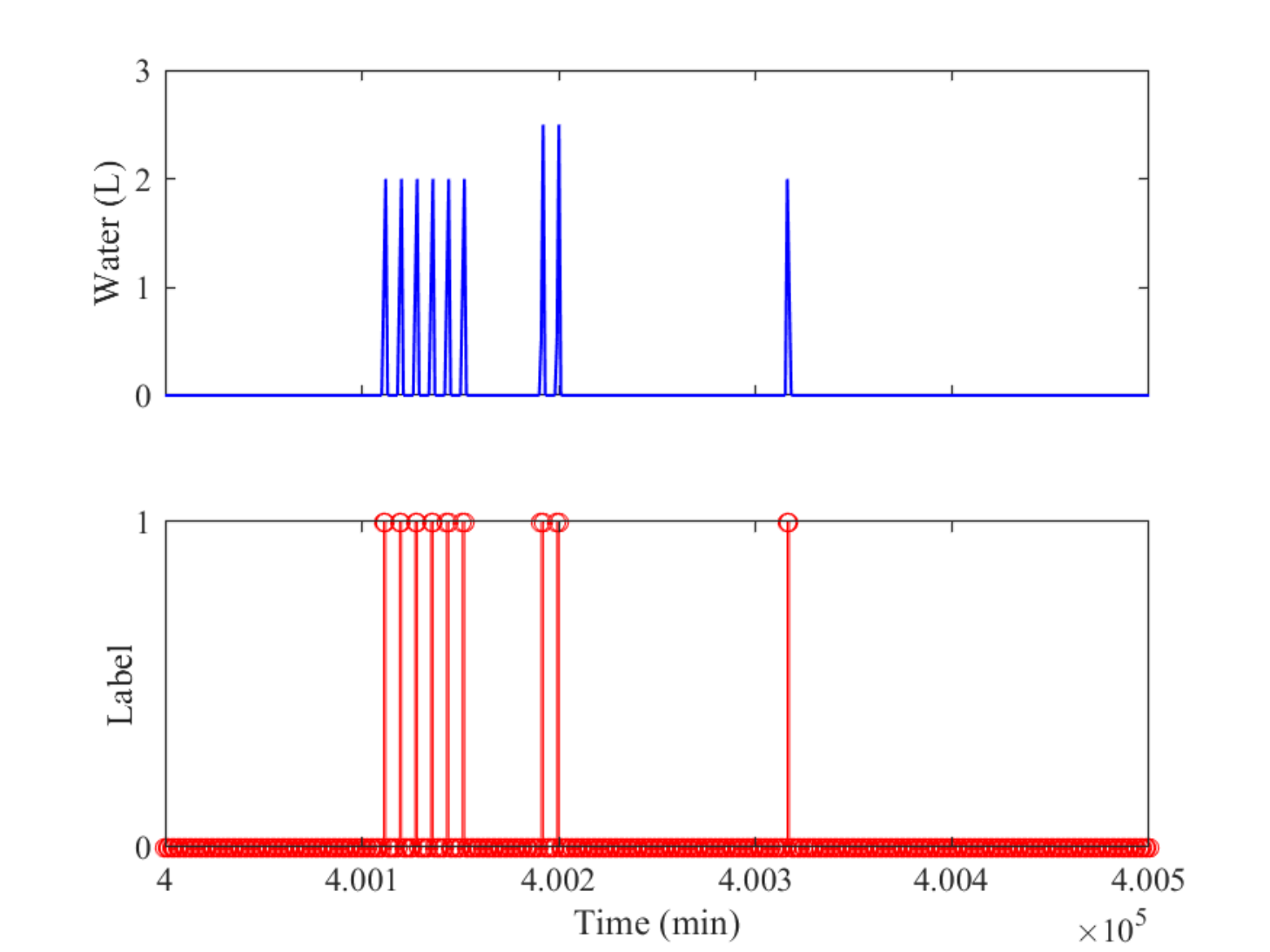}\vspace{.05in}
        \caption{Dishwasher's water signal and corresponding labels}
        \label{water_recons}
    \end{figure}
\end{enumerate}

%%%%%%%%%%%%%%%%%%
%%%%%%%%%%%%%%%%%%
\section{Proposed Classification-based NILM Method}

    In this section, we propose an event-based classification NILM method that is highly accurate and has reasonably a low computational complexity. This algorithm utilizes the KNN algorithm in the first phase to extract the power signals of appliances that do not have overlapping power values with any other appliance. Then, to take advantage of the correlations between water and power signals of some appliances, %and the capability of deep learning-based methods in learning hidden features without any manual effort,
    two novel deep learning algorithms are proposed, specifically aiming to extract the power signals of appliances with close power values. In each of these methods, water consumption is considered as a new signature for the NILM purpose.
\subsection{Phase 1: KNN Classification}
    
    In this phase, the KNN classification algorithm, which is simple and fast, is utilized for extracting the consumption profiles of appliances with exclusive power values. The model is trained using the event signal constructed in Section~\ref{preprocessing} and the corresponding labels of its events.

%%%%%%%%%%%%%%%%%%
\subsection{Phase 2: LSTM-based Classification}\label{phase 2}

\begin{figure}[t]
        \centering
        \includegraphics[width=.5\linewidth]{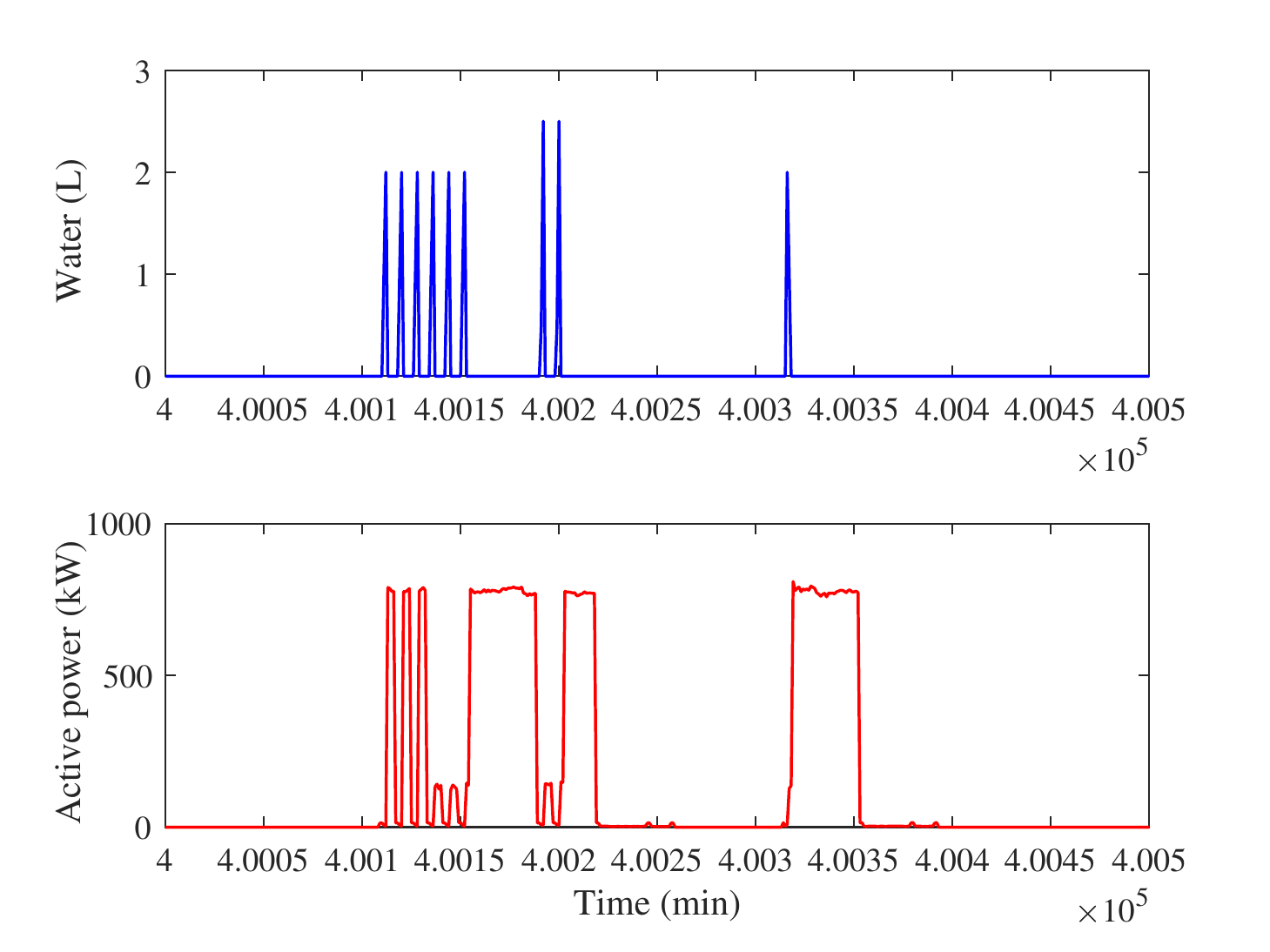}
        \caption{Water and power signals of the dishwasher}
        \label{DW_water_power}
    \end{figure}

    There is correlation between power and water consumption in any residential building due to the following reasons. First, there are appliances in the house which consume both electricity and water. For instance, the dishwasher has a specific electricity consumption pattern or program which consists of washing, rinsing and drying \cite{kong2016extensible}, and at each part of this pattern, specific amounts of water are consumed. Second, some activities of residents lead to simultaneous, and correlated, power and water consumption. As an example, anytime the bathroom is in use, its lights are on while water is being consumed. We will focus on the first reason stated and in particular take advantage of correlation between the power and water signals of the dishwasher, the existence of which should be evident in Fig.~\ref{DW_water_power}, in order to improve the NILM accuracy. Two separate algorithms capable of exploiting said correlation are proposed below.

%%%%%%%%%%%%%%%%%%
\subsubsection{Parallel Disaggregation of Power and Water Signals}

\begin{figure*}[t]
        \centering
        \includegraphics[width=\linewidth]{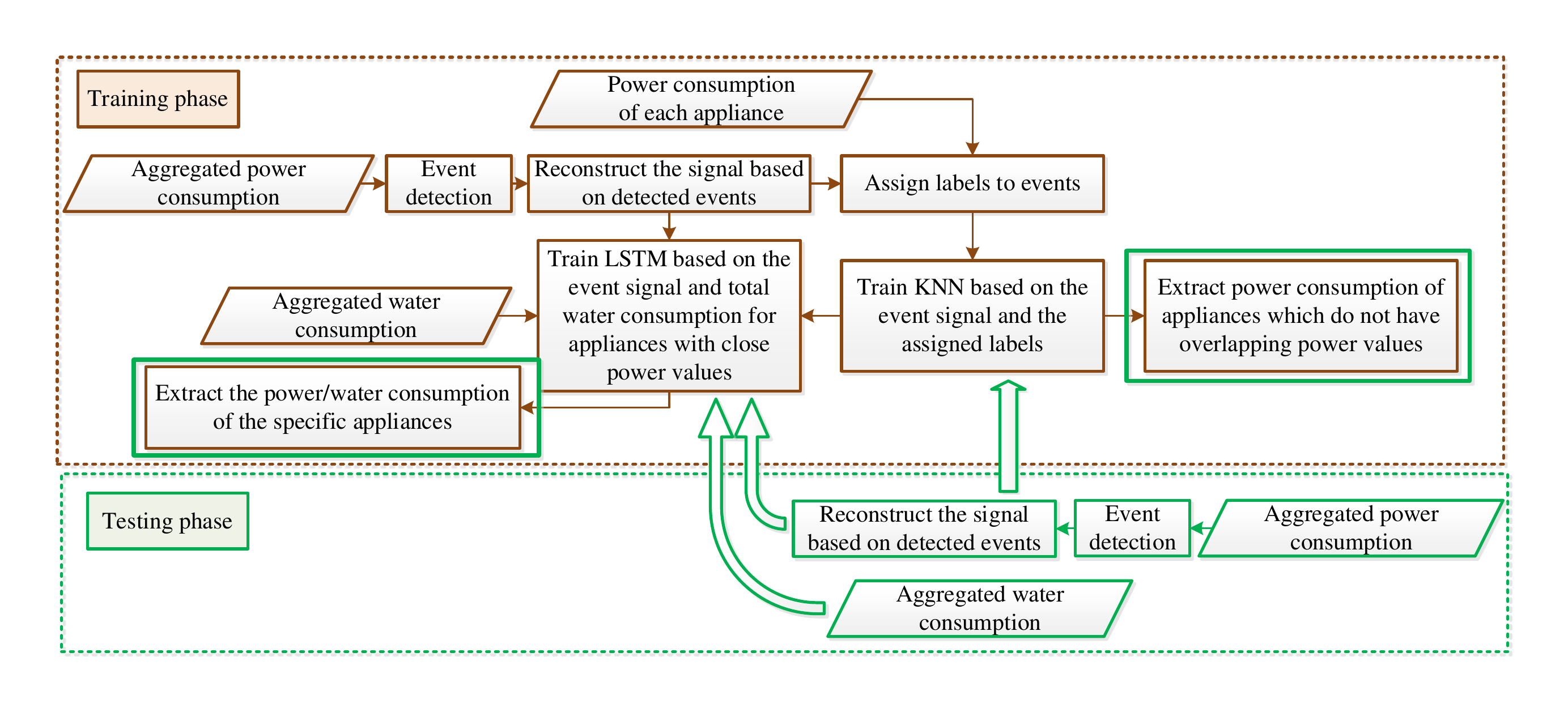}
        \caption{The flowchart of the proposed NILM process with parallel disaggregation}
        \label{Syn}
    \end{figure*}

    Having balanced the data, for each appliance consuming both electricity and water, an LSTM network is designed. The event signal and the aggregated water signal form the input of the network, while the assigned {\color{black}power and water} labels of that appliance are considered as the output. We point out that appliances with overlapping power values, such as the dishwasher, are of particular interest. This model uses correlation between the aggregated power and water signals to disaggregate both signals in parallel. Fig.~\ref{Syn} details a complete NILM process, Phase 1 of which consists of the KNN classification algorithm, while Phase 2 of which consists of the algorithm presented here that carries out parallel disaggregation of the power and water signals.

%%%%%%%%%%%%%%%%%%
\subsubsection{Iterative Disaggregation of Power and Water Signals}
 
\begin{figure*}[t]
        \centering
        \includegraphics[width=\linewidth]{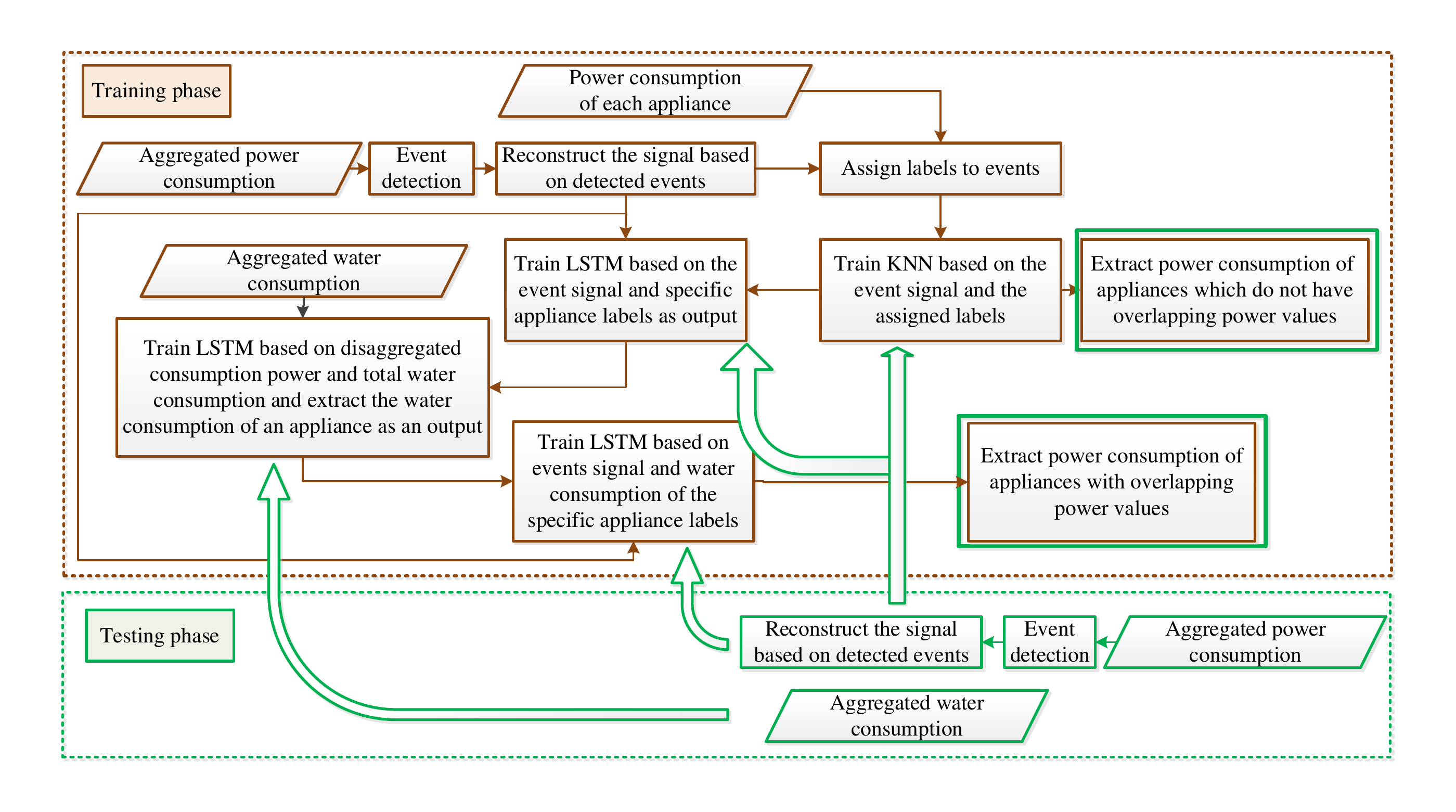}
        \caption{The flowchart of the proposed NILM process with iterative disaggregation}
        \label{Asy}
    \end{figure*}

    In this scheme, three serial LSTM networks are designed for every appliance that consumes both electricity and water. The input of the first LSTM is the {\color{black}aggregated power signal} and its output is the power signal of the appliance. Then, this output along with the aggregated water signal is considered as the input to the second LSTM network, while the water signal of that specific appliance is considered as its output. Finally, the last network is trained considering the {\color{black}aggregated power signal} and the output of the second network as inputs and the power signal of the specific appliance as the output. In other words, in this method, the extracted power signal of the  appliance improves the accuracy of extracting its water signal, then the extracted water signal improves the power disaggregation accuracy. Fig.~\ref{Asy} demonstrates the flowchart of a complete NILM process, where the KNN algorithm is used in Phase 1, while Phase 2 consists of the algorithm presented here for iterative disaggregation of power and water signals. 
%%%%%%%%%%%%%%%%%%
\section{Numerical Results}

    In this section, the NILM processes depicted in Fig.~\ref{Syn} and \ref{Asy} are examined via the well-known AMPds dataset \cite{makonin2013ampds}. A comparison between these results and those of a method proposed recently in the literature will be presented in the last subsection.
    
    In the following subsections, the features of AMPds and evaluation metrics will be discussed. Then, after applying pre-processing algorithm on the power and water signal, the proposed parallel and iterative disaggregation methods will be applied. Finally, the results will be compared with existing method in this field.

%%%%%%%%%%%%%%%%%%
\subsection{AMPds Dataset}
    
\begin{figure}[t]
       \centering
        \includegraphics[width=.5\linewidth]{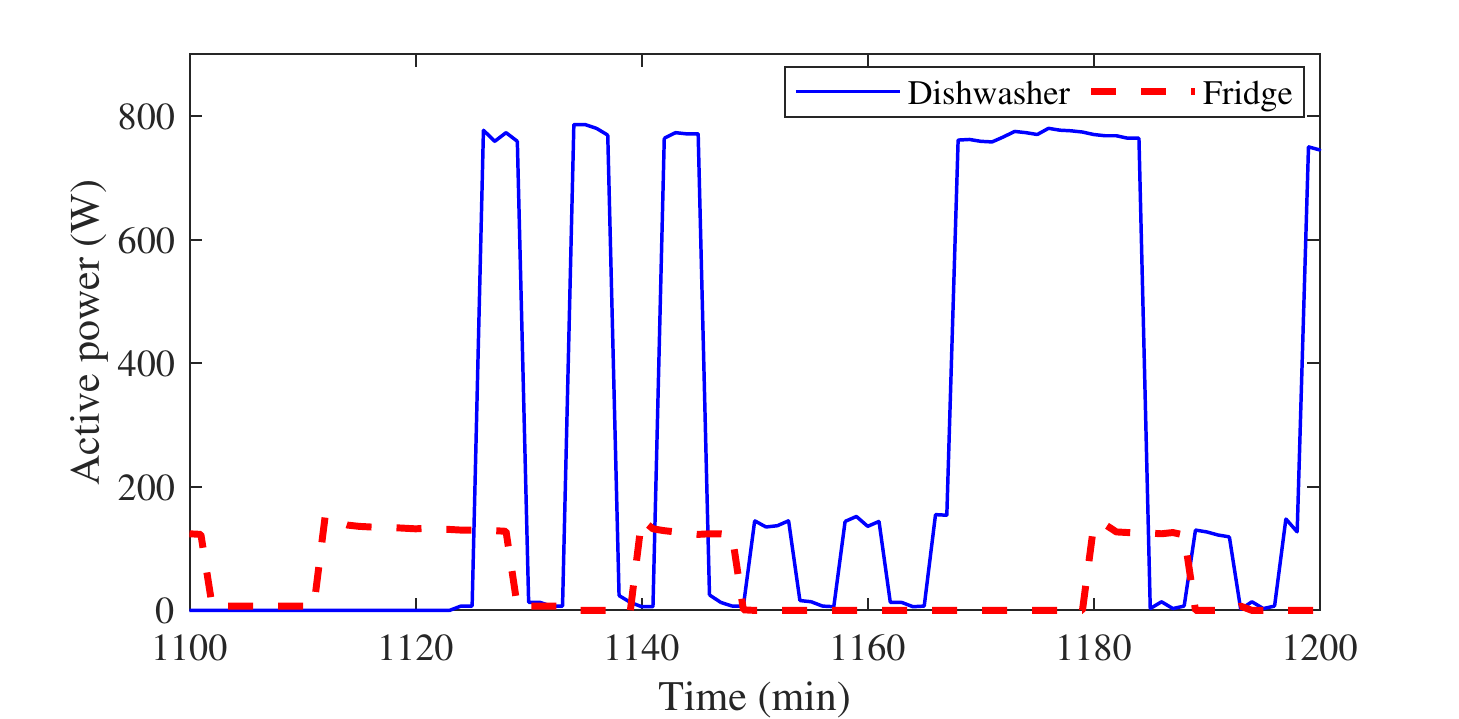}\vspace{.1in}
        \caption{Power signals of the fridge and dishwasher}
        \label{Dishwasher_fridge}
    \end{figure}

\begin{figure}[t]
        \centering
        \includegraphics[width=.5\linewidth]{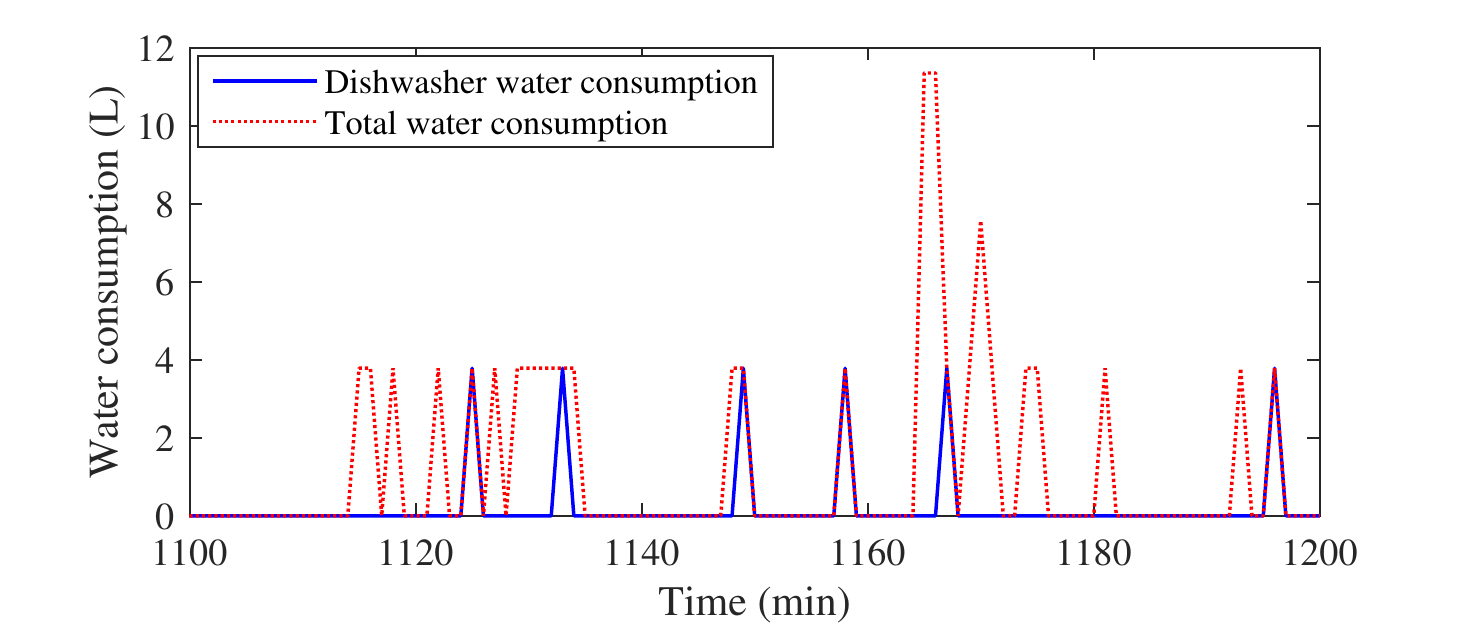}\vspace{.1in}
        \caption{Aggregated and dishwasher's water signals}
        \label{DW_total}
    \end{figure}

    The AMPds dataset includes a total of 1,051,200 minutely readings from 21 power meters, 2 water meters, and 2 natural gas meters of one residence in a span of 730 consecutive days \cite{makonin2013ampds}. We use 550 days of data as the training dataset and 180 days of data as the test dataset. 7 different appliances, including the fridge and dishwasher as multi-mode appliances, are considered to evaluate the performance of the proposed methods. The 5 other appliances are the dryer, basement electricity, oven, washing machine and heat pump, which are all used frequently. Operation modes of these appliances and their corresponding power values are extracted based on the authors’ earlier work in \cite{azizi2020residential} and listed in Table~\ref{modes}. As it is shown in Fig.~\ref{Dishwasher_fridge}, the fridge and dishwasher have overlapping power values, which can reduce the accuracy of classification. The aggregated water signal is depicted in Fig.~\ref{DW_total} along with that of the dishwasher.
    \begin{table}[H] \centering
        \caption{Extracted power values at each appliance mode based on mode extraction method in \cite{azizi2020residential}}
        \begin{tabular}{|c|c|c|c|}
            \hline
            Appliance	& Mode 1	& Mode 2	& Mode 3 \\ \hline
            Fridge &	0	&100-200&	400-500 \\ \hline
            Dryer&	0&	4000-5000&	- \\ \hline
            Dishwasher&	0	&100-200&	700-800\\ \hline
            Heat pump&	0	&1000-1850	&-  \\ \hline
            {\color{black}Oven} & 0 & 3400-3550 &- \\ \hline
            {\color{black}Basement} & 0 & 330-350 & - \\ \hline
              {\color{black}Washing machine} & 0 & {\color{black}100-250} & 400-700\\ \hline
        \end{tabular} \label{modes}
    \end{table}

%%%%%%%%%%%%%%%%%%
\subsection{Evaluation Metrics}

    Evaluation metrics used in this work are F-measure, precision, and recall 
    \cite{faustine2017survey}, that are calculated according to the following equations:
    \begin{equation}
        F_{measure} = \frac{2\times TRP \times RC}{TRP+RC},
	    \label{Acc2}
    \end{equation}
    \begin{equation}
        Precision = \frac{TP}{TP+FN},
	    \label{Acc3}
    \end{equation}
    \begin{equation}
        Recall = \frac{FP}{FP+TN},
	    \label{Acc4}
    \end{equation}
    where $TP$ (true positive) and $TN$ (true negative) denote the number of correct diagnoses of the ON and OFF modes, respectively, while $FP$ (false positive) and $FN$ (false negative) are the numbers of incorrect diagnoses of the OFF and ON modes, respectively. The F-measure metric which is the harmonic average of precision and recall, is widely used in the literature to evaluate the efficiency of NILM methods. Based on this metric, multi-mode appliances are treated as ON/OFF appliances all their non-OFF modes are considered as one ON mode.
    
    %determine the efficacy of the proposed method to NILM.

%%%%%%%%%%%%%%%%%%
\subsection{Pre-processing}

    In the pre-processing stage of the two proposed NILM processes, the event signal, the events' labels, and the labels of each sample of the aggregated water signal are derived based on the proposed methods in Section \ref{preprocessing}. We perform class balancing via under-sampling that will render the NILM processes much more time-efficient than over-sampling \cite{de2019boosting}, and yet achieve strikingly accurate results.

\subsection{Evaluating the Proposed NILM Processes}

    To extract the contribution of each appliance in aggregated consumption, in Phase 1, the KNN algorithm with $K=7$ and Minkowski distance metric is employed. The F-measure of detecting each appliance is presented in Table~\ref{KNN_acc}. As expected, this algorithm is able to accurately detect appliances with exclusive power values, namely the dryer, oven, basement electricity, and heat pump, in the aggregated power signal. However, its performance is not as good for the other three appliances, namely the fridge, washing machine, and dishwasher, that have overlapping power values. To compensate for the relatively poor accuracy with regard to the fridge and dishwasher, Phase 2 of our NILM processes, described in Subsection \ref{phase 2}, is then carried out. We separately discuss and evaluate both algorithms presented in Subsection \ref{phase 2} below. The values for the parameters of the LSTM networks are listed in Table~\ref{P}.
    
    Using the parallel power and water disaggregation leads to a significant improvement in the accuracy of detecting the fridge and dishwasher, with the results presented in Table~\ref{T1}. Moreover, Table~\ref{T2} shows the accuracy of the dishwasher's water disaggregation. Finally, the use of the iterative power and water disaggregation algorithm leads to the nearly perfect results, given in Table~\ref{T3} for the power disaggregation of the fridge and dishwasher and in Table~\ref{T4} for the dishwasher's water disaggregation.
    
    %To disaggregate the power consumption of appliances with close consumption power and LSTM is utilized. All results are obtained considering the neural network parameters of illustrated in Table \ref{P} for the model training. The total water consumption is considered as a new signature inparallel method and the disaggregated water of dishwasher is utilized in itterative method which will be discussed in the following.

    \begin{table}[H]\centering
        \caption{Accuracy evaluation for the KNN algorithm}
        \begin{tabular}{|c|c|c|c|}
            \hline
            Appliance & $Precision$ & $Recall$ & $F_{measure}$ \\ \hline
            Fridge &  0.73 & 0.82 & 0.77 \\ \hline
            Dishwasher & 0.74 & 0.77 & 0.76 \\ \hline
            Dryer & 0.95 & 0.98 & 0.97 \\ \hline
            Heat pump & 0.96 & 0.97 &  0.96\\ \hline
            {\color{black}Basement} &0.93 &0.98 &0.95\\ \hline
            {\color{black}Clothes washer} &0.84 & 0.65&0.73\\ \hline
            {\color{black}Oven} &0.99 & 0.95&0.97\\ \hline
        \end{tabular} \label{KNN_acc}
    \end{table}%\vspace{-.05in}
    
    \begin{table}[H] \centering
        \caption{Model training parameters}
        \begin{tabular}{|c|c|}
            \hline
            Item & Value \\ \hline
            Activation function & Relu, Soft max \\ \hline
            Dropout probability & 0.3 \\ \hline
            Mini-batch size & 50 \\ \hline
            Number of epochs & 50 \\ \hline
            Optimizer & Adam \\ \hline
            Learning rate & 0.001 \\ \hline
            Data pre-processing & Min-Max scaler \\ \hline
            Loss function & Binary cross-entropy \\ \hline
            Metrics & Accuracy \\ \hline
            LSTM layer size & 500 \\ \hline
            Fully connected layer size & 200 \\ \hline
            Kernel initializer & Random uniform \\ \hline
        \end{tabular} \label{P}
    \end{table}%\vspace{-.05in}
%\subsection{Water and power consumption of the dishwasher}
%Analyzing total water consumption of dishwasher as shown in Fig. \ref{figDW1} obviously, there is a specific pattern include of dishwasher watershed periods, similar values in each period, and time intervals that would provide a good model for extracting dishwasher water consumption from total water consumption. Moreover, It is clear from Fig. \ref{figDW2} each event from OFF mode to ON mode aligns with dishwasher watershed. These phenomena provide a favorable condition for reaching appropriate accuracy in the disaggregation problem.
%\begin{figure}
%    \centering
%    \includegraphics[scale=0.55]{Water_DW_Total.eps}
%    \caption{Total water consumption vs dishwasher's water consumption}
%    \label{figDW1}
%\end{figure}

    \begin{table}[H]\centering 
        \caption{Accuracy evaluation of power disaggregation for the parallel disaggregation algorithm}
        \begin{tabular}{|c|c|c|c|}
            \hline
            &$Precision$ & $Recall$ & $F_{measure}$  \\ \hline
            Fridge &0.94&0.99&0.96  \\ \hline
            Dishwasher &0.99&0.94&0.96  \\ \hline
        \end{tabular}
    \label{T1}
    \end{table}%\vspace{-.05in}
    
    \begin{table}[H]\centering
        \caption{Accuracy evaluation of water disaggregation for the parallel disaggregation algorithm}
        \begin{tabular}{|c|c|c|c|}
            \hline
            & $Precision$ & $Recall$ & $F_{measure}$  \\ \hline
%Total water consumption &0.92&0.92&0.92 \\ \hline
            Dishwasher water consumption &0.92&0.92&0.92 \\ \hline
        \end{tabular}
    \label{T2}
    \end{table}%\vspace{-.05in}
    
    \begin{table}[H]\centering
        \caption{Accuracy evaluation of power disaggregation for the iterative disaggregation algorithm}
        \begin{tabular}{|c|c|c|c|}
            \hline
            & $Precision$ & $Recall$ & $F_{measure}$  \\ \hline
            Fridge &0.99&0.99&0.99  \\ \hline
            Dishwasher &0.99&0.99&0.99 \\ \hline
        \end{tabular} \label{T3}
    \end{table}%\vspace{-.05in}
    
    \begin{table}[H]\centering
        \caption{Accuracy evaluation of water disaggregation for the iterative disaggregation algorithm}
        \begin{tabular}{|c|c|c|c|} 
            \hline
            & $Precision$ & $Recall$ & $F_{measure}$  \\ \hline
%Total water consumption &0.92&0.93&0.93 \\ \hline
            Dishwasher water consumption &0.93&0.92&0.93\\ \hline
        \end{tabular}\label{T4}
    \end{table}
%%%%%%%%%%%%%%%%%%
\subsection{Comparison and Discussion}

    In this subsection, we further discuss the classification results of our proposed NILM processes and compare them with those of an existing method in the literature.
    
\subsubsection{Comparison with individual LSTM networks for power and water disaggregation}

    To evaluate the efficiency of the proposed parallel and iterative disaggregation algorithms, we consider two separate LSTM networks for power and water disaggregation considering the power and water signals as their respective inputs. We then report validation accuracy to test the performance of these two individual networks versus that of our proposed algorithms in encountering data which it has not been seen in the past in Table~\ref{val_1} and \ref{val_2}. As it is shown, our proposed algorithms perform better than their counterparts that do not take advantage of correlation between the power and water signals.

\subsubsection{Comparison with another classification method}  
    
    We now compare our results with those of the entropy-based classification method in \cite{machlev2018modified}, as reported comprehensively in Table~\ref{comparison}. It can be seen that utilizing power and water consumption in parallel and iterative manners increases the overall accuracy of disaggregation even considering higher number of modes ($N_{modes}$) for appliances. %Furthermore, as it is displayed in Table~\ref{comparison}, in comparison with \cite{machlev2018modified} we considered more operation modes for multi-mode appliances.}

\begin{table}[H]\centering
\caption{Comparing performance of different methods in power disaggregation}
\begin{tabular}{|c|c|c|c|}
\hline
 Method & Validation accuracy   \\ \hline
\begin{tabular}[c]{@{}c@{}}Power disaggregation with an LSTM \end{tabular} & 0.95  \\ \hline
\begin{tabular}[c]{@{}c@{}}Power disaggregation with parallel method \end{tabular} 
& 0.96  \\ \hline
\begin{tabular}[c]{@{}c@{}}Power disaggregation with iterative method  \end{tabular} 
& 0.99  \\ \hline

\end{tabular} \label{val_1}
\end{table}

\begin{table}[H]\centering
\caption{Comparing performance of different methods in water disaggregation}
\begin{tabular}{|c|c|c|c|}
\hline
Method & Validation accuracy   \\ \hline
\begin{tabular}[c]{@{}c@{}}Water disaggregation by an LSTM\end{tabular} & 0.81  \\ \hline
\begin{tabular}[c]{@{}c@{}}Water disaggregation with parallel method \end{tabular}
 &0.92  \\ \hline
\begin{tabular}[c]{@{}c@{}}Water disaggregation with iterative method  \end{tabular}
  &0.93  \\ \hline

\end{tabular} \label{val_2}
\end{table}

% \usepackage{multirow}
% \usepackage{multirow}
\begin{table}[H]\centering
\caption{Comparison of F-measure for different methods}
\begin{tabular}{|c|c|c|c|c|c|}
\hline
\multirow{3}{*}{Apps}                                            & \multicolumn{3}{c|}{Proposed method}                                                                 & \multicolumn{2}{c|}{\multirow{2}{*}{\cite{machlev2018modified}}}                               \\ \cline{2-4}
                                                                      & \multirow{2}{*}{\begin{tabular}[c]{@{}c@{}}$N_{modes}$\end{tabular}} & Parallel  & Iterative  & \multicolumn{2}{c|}{}                                                   \\ \cline{1-1} \cline{3-6} 
\multicolumn{1}{|l|}{}                                                &                                                                             & F-measure & F\_measure & \begin{tabular}[c]{@{}c@{}}$N_{modes}$\end{tabular} & F\_measure \\ \hline
HP                                                                    & 3                                                                           & 0.96      &  0.96          & 1                                                          & 0.97       \\ \hline
DW                                                                  & 2                                                                           & 0.96      &    0.99        & 1                                                          & 0.72       \\ \hline
CW                                                                   & 3                                                                           & 0.73      &     0.73       & -                                                          & -          \\ \hline
FRG                                                                   & 2                                                                           & 0.96      &     0.99       & 1                                                          & 0.88       \\ \hline
BM                                                                   & 1                                                                           & 0.95      &      0.95      & 1                                                          & 0.86       \\ \hline
FT                                                                & 1                                                                           & -      &    -        & 1                                                          & 0.93       \\ \hline
OVN                                                               & 1                                                                           & 0.97      &   0.97         & 1                                                          & 0.50       \\ \hline
Dryer                                                                 & 1                                                                           & 0.97         &   0.97         & 3                                                          & 0.29       \\ \hline
\textbf{AVG}  &                                                                             & 0.932      &    0.937        & \multicolumn{2}{c|}{0.735}                                               \\ \hline
\end{tabular}
\label{comparison}
\end{table}

%%%%%%%%%%%%%%%%%%
%%%%%%%%%%%%%%%%%%
\section{Conclusion}

    In this paper, we have proposed novel, practical classification processes for the NILM problem. The processes are tailored in such a way to address a well-known NILM challenge, that is the existence of appliances with overlapping power values. The proposed NILM processes take advantage of correlation between the aggregated power and water signals of a residence to achieve high accuracy. Each process consists of two main phases. In the first phase, that is shared between the two processes, the KNN classification algorithm is employed to disaggregate the power consumption of appliances with exclusive power values. Phase 2 of each process involves a deep learning-based method, where one carries out parallel disaggregation of the power and water signals, while the other carries out disaggregation of the two signals in an iterative fashion.
    %parallel and iterative methods, considering the water consumption as a novel feature in the LSTM network are proposed. In the parallel method, power and water consumption are disaggregated in a parallel way. However, in an iterative algorithm, the disaggregated power signal improves the accuracy of the dishwasher's water disaggregation and its disaggregated water improves the accuracy of power disaggregation.
    The proposed NILM processes have been evaluated considering 7 different appliances of the AMPds dataset, where their strong performance have been confirmed.

    The main idea for future work is to capitalize on parallel disaggregation of energy, by having gas consumption involved in the process. We believe that there is clear correlation between the aggregated power, water, and gas signals of a residential house that should be exploited. 
%    \\a few lines\\a few lines\\a few lines\\a few lines\\a few lines\\a few lines
    %Then, we will continue to extract the water/gas consumption of some appliances considering the disaggregated power and total water/gas consumption which omits the requirement for the training dataset and is more practical in the real world.

%%%%%%%%%%%%%%%%%%
%%%%%%%%%%%%%%%%%%
\balance
\bibliography{Ref}
\bibliographystyle{IEEEtran}
\noindent

%%%%%%%%%%%%%%%%%%
%%%%%%%%%%%%%%%%%%
\end{document}